\documentclass[11pt]{article}
\usepackage{moriond}
\usepackage{epstopdf}
\def\Journal#1#2#3#4{{#1} {\bf #2}, #3 (#4)}

\def\PLB{{\em Phys. Lett.}  B}

\def\PRC{{\em Phys. Rev.} C}

\def\JHEP{{\em JHEP}}

\def\be{\begin{equation}}
\def\ee{\end{equation}}
\def\bea{\begin{eqnarray}}
\def\eea{\end{eqnarray}}

\def\jpsi{J$/\psi$}
\def\ro{$\rho^0$ }
\def\pt{$p_{\rm T} $}
\def\etal{{\it et al}}

\newcommand{\Photo}{}

\begin{document}
\vspace*{4cm}
\title{DIFFRACTION AND ULTRAPERIPHERAL COLLISIONS AT ALICE}

\author{ E. L. Kryshen for the ALICE Collaboration}

\address{
Petersburg Nuclear Physics Institute, National Research Center ''Kurchatov Institute"\\
1 Orlova Roscha, 188300 Gatchina, Russia}

\maketitle\abstracts{
The ALICE experiment measured \jpsi{} photoproduction in Pb+Pb ultraperipheral collisions at $\sqrt{s_{NN}}=2.76$ TeV at forward and central rapidity. 
The coherent \jpsi{} cross section is found to be in good agreement with those models which include nuclear gluon shadowing consistent with EPS09 parametrization. 
Results on \ro photoproduction in ultra-peripheral Pb+Pb collisions are
also briefly discussed. Studies on exclusive \jpsi{} in p+Pb and on diffraction in pp collisions are mentioned.
}
\section{Introduction}

Ultraperipheral collisions (UPC) are characterised by impact parameters larger than two nuclear radii. In this case hadronic interactions are strongly suppressed resulting in a dominance of electromagnetic processes caused by large flux of quasi-real photons from heavy nuclei. 

Vector meson photoproduction on nuclear target ($\gamma A \to V A$) is one of the most interesting processes which can be studied in heavy ion UPC~\cite{review}. One has to distinguish between coherent and incoherent photoproduction. In the first case, photon couples coherently to the whole nucleus causing a narrow transverse momentum distribution of the produced meson (mean $p_{\rm T} \sim 60$~MeV$/c$). The target nucleus stays intact in about 80\% of coherent events. In the incoherent case, the photon couples to a single nucleon in the target, thus vector meson \pt{ }distribution is dictated by the nucleon form factor and becomes much broader (mean $p_{\rm T} \sim 400$~MeV$/c$). The incoherent process is usually accompanied by neutron emission from the target.

The coherent heavy quarkonium photoproduction is of particular interest since, in the leading order pertubative QCD, its cross section is proportional to the squared gluon density in the target. Thus the measurement of the coherent \jpsi{} photoproduction in Pb+Pb UPC provides a direct tool to study poorly known nuclear gluon shadowing effects which play crucial role in the calculation of initial state parton distributions in heavy ion collisions.

The ALICE experiment perfectly matches all requirements for such a measurement. On the one hand, the UPC analysis strategy relies on the selection of events with only two leptons from \jpsi{} decay and otherwise empty detector, therefore large continuous angular coverage in ALICE is essential to ensure event emptiness. On the other hand, ALICE can trigger on \jpsi{} down to zero transverse momentum which is important for the measurement of the low-$p_{\rm T}$ coherent \jpsi{} photoproduction. Further details on the ALICE experimental setup can be found in~\cite{alice}.

\section{J$/\psi$ photoproduction}

ALICE measured \jpsi{} photoproduction in Pb+Pb UPC both at forward and central rapidity.

The forward \jpsi{} measurement was performed with the muon spectrometer and is based on 2011 Pb+Pb data at $\sqrt{s_{NN}} = 2.76$ TeV. The analysis was recently published in Ref.~\cite{coherent-forward}.
The trigger required a single muon with \pt$ > 1$ GeV$/c$ in the muon arm acceptance, at least one cell fired in VZERO-C scintillator array overlapping with the muon arm and veto on VZERO-A activity on the opposite side. Event emptiness was ensured by vetoing large activity in the neutron zero-degree-calorimeters and the silicon pixel detector at central rapidity.  

\begin{figure}
\includegraphics[width=0.49\linewidth]{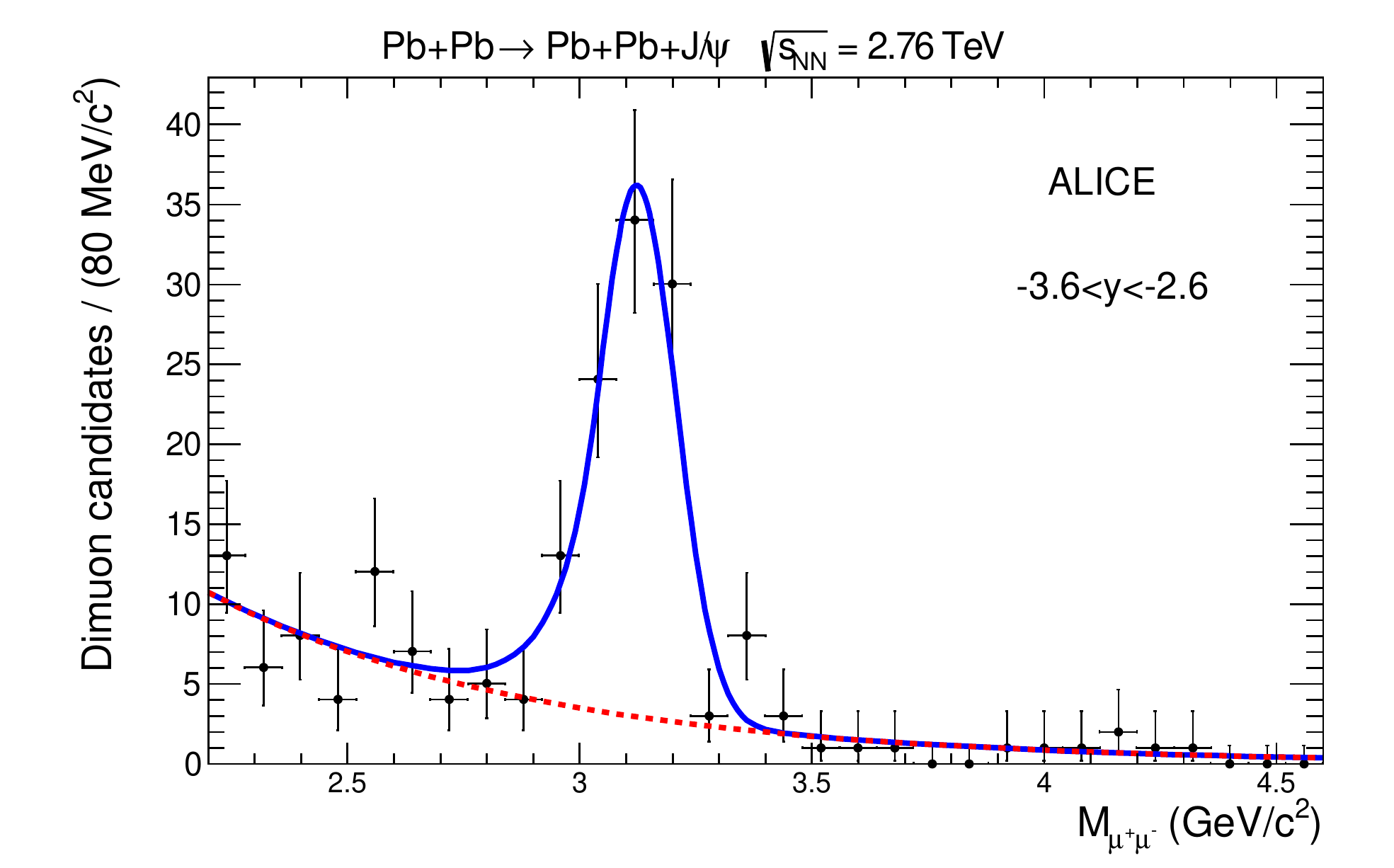}
\includegraphics[width=0.49\linewidth]{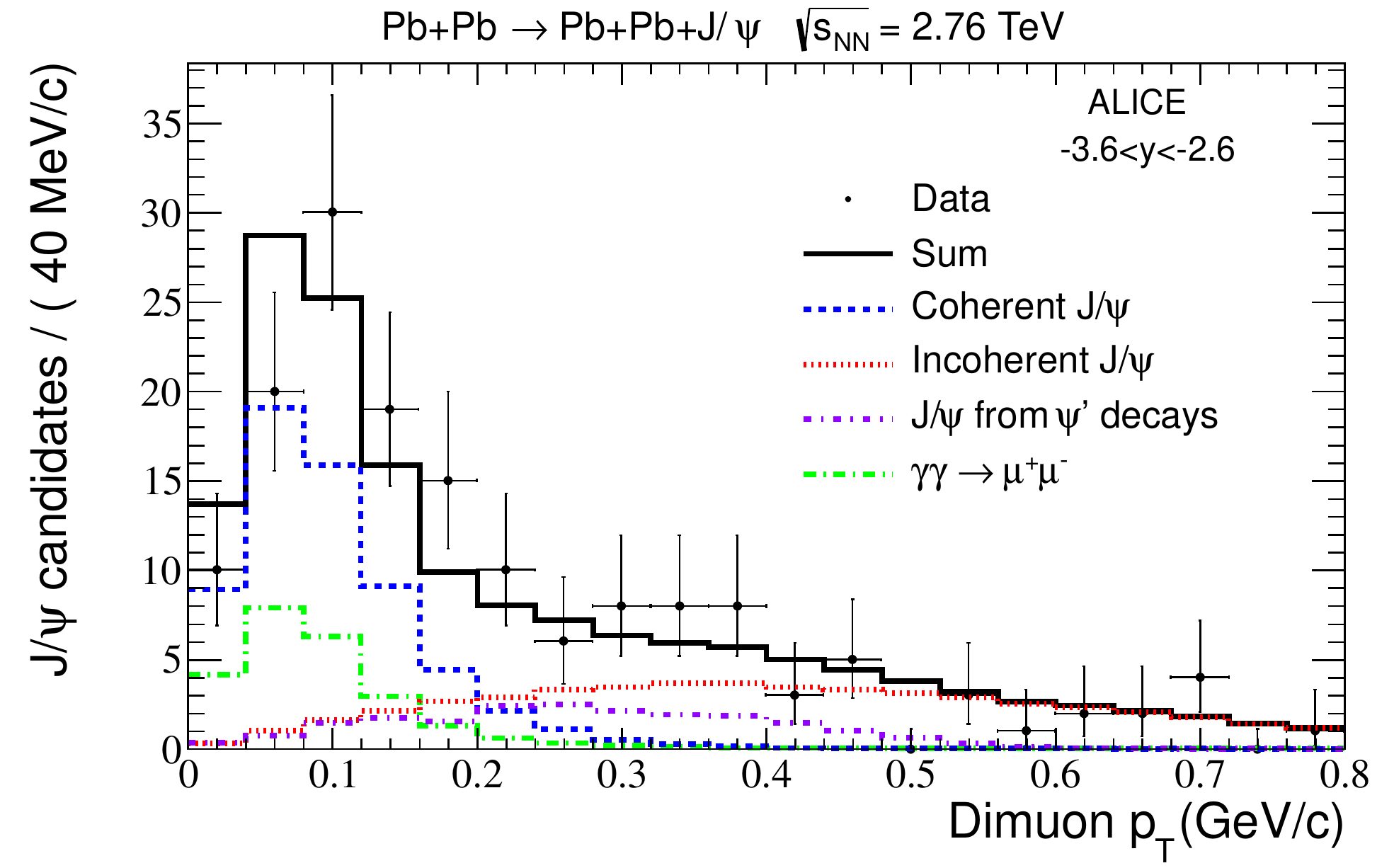}
\caption{Dimuon invariant mass spectrum with $p_{\rm T} <300$ MeV$/c$ (left) and transverse momentum distribution for dimuons in the invariant mass range $2.8 {\ \rm GeV}/c^2 < M_{\mu^+\mu^-} < 3.4 {\ \rm GeV}/c^2$ (right). From Ref.~\protect\cite{coherent-forward}}
\label{fig:forward_jpsi}
\end{figure}

The resulting invariant mass spectrum for opposite-sign dimuons with $p_{\rm T} <300$ MeV$/c$ in the rapidity range $-3.6 <y < -2.6$ is shown in fig.~\ref{fig:forward_jpsi} (left). The signal was fitted with the Crystal Ball function on top of irreducible background fitted with an exponential shape. The background shape appeared to be in good agreement with expectations from the continuum $\gamma\gamma\to\mu^+\mu^-$ production.

The transverse momentum distribution for opposite-sign dimuons in the invariant mass range $2.8 {\ \rm GeV}/c^2 < M_{\mu^+\mu^-} < 3.4 {\ \rm GeV}/c^2$ is shown in fig.~\ref{fig:forward_jpsi} (right). This spectrum was fitted with a sum of Monte-Carlo templates corresponding to four contributions: coherent and incoherent \jpsi, feed-down \jpsi{} from $\psi'$ decays and $\gamma\gamma \to \mu^+\mu^-$ contribution at low \pt. The coherent \jpsi{} yield was extracted from the fit keeping relative contributions of coherent and incoherent \jpsi{} unconstrained.

Normalization of the coherent \jpsi{} cross section at forward rapidity was performed with respect to the continuum dimuon pair production cross section which is known from QED but with large uncertainties of about 20\%. The cross section $d\sigma^{\rm coh}_{J/\psi}/dy =  1.00 \pm 0.18\, {\rm (stat)} {}^{+0.24}_{-0.26}\, {\rm (syst)}$~mb in the rapidity interval $-3.6 <y < -2.6$ has been recently published by ALICE~\cite{coherent-forward}.

The measurement of \jpsi{} production in UPC at midrapidity was 
obtained using data collected in 2011.
The trigger required back-to-back topology of the fired cells in the time-of-flight detector (TOF), at least two hits in the silicon pixel detector (SPD) and vetoes on VZERO counters at forward and backward rapidity. Track reconstruction was performed with the time projection chamber (TPC) and the silicon tracking system in about two pseudorapidity units. 
Events with only two opposite sign good-quality tracks coming from a recontructed primary vertex have been selected.
\jpsi{} signal was extracted both in dielectron and dimuon channels which were separated by the energy deposition in TPC. Coherent and incoherent contributions in the \jpsi{} peak were extracted from fits to dilepton $p_{\rm T}$ spectra similar to the forward case. Cross section normalization was performed with respect to the integrated luminosity which was measured using a trigger for the most central hadronic Pb+Pb collisions and corrected for the UPC trigger live time. Results have been recently published in~\cite{coherent-central}.

\begin{figure}
\centering
 \includegraphics[width=0.6\linewidth]{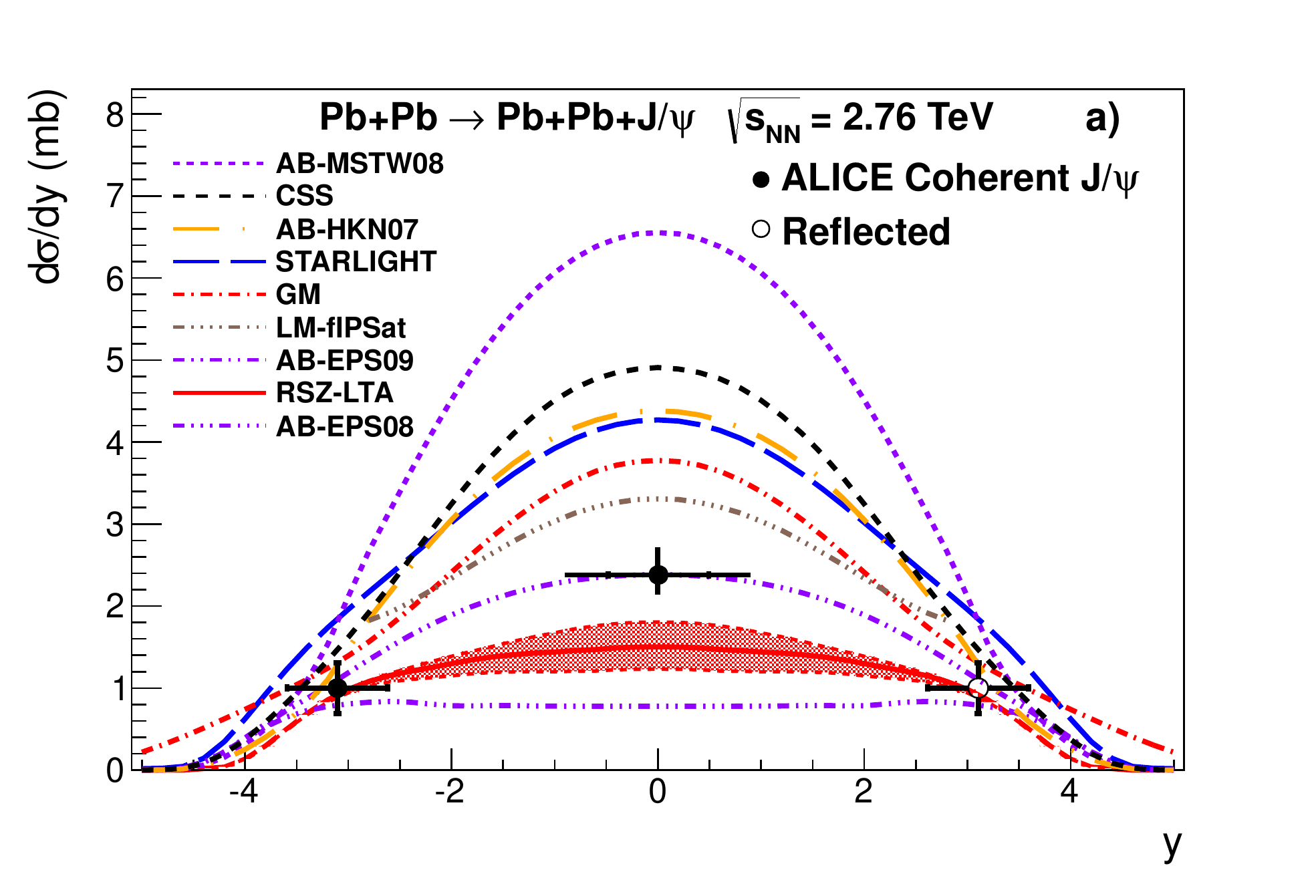}
\caption{
ALICE results on the coherent \jpsi{} photoproduction cross section at forward and central rapidity 
compared with model predictions. See Ref.~\protect\cite{coherent-forward,coherent-central} for details.}
\label{fig:cs}
\end{figure}

ALICE results on coherent \jpsi{} photoproduction cross section at forward ($-3.6 <y < -2.6$) and central ($|y| < 0.9$) rapidity  are compared with various model calculations in fig.~\ref{fig:cs}. There are at least 5 predictions which have been recently published. The STARLIGHT generator is based on the Vector Dominance Model coupled with the classical probabilistic formula to account for \jpsi{} absorption effects~\cite{sl}. A set of predictions by Adeluyi--Bertulani (AB) is based on LO pQCD calculations: the upper curve AB-MSTW08 assumes no modification of the gluon distribution while other three incorporate gluon shadowing from various parameterizations (HKN07, EPS08, EPS09)~\cite{ab}. 
The difference of the measured cross section with respect to the upper curve reflects the strength of the gluon shadowing effect. Goncalves-Mochado (GM) prediction is based on the color dipole model and saturation effects from the Color-Glass-Condensate framework~\cite{gm}.  CSS model accounts for shadowing effects by propagating $c\bar c$ and $c\bar c g$ intermediate states in the Glauber approach~\cite{cisek}. Finally, RSZ-LTA model incorporates gluon shadowing effects in the leading twist approximation~\cite{lta}.

The obtained experimental data favour the models which take into account gluon shadowing effects in the LO pQCD framework.  The best agreement is found for the model~\cite{ab} which incorporates gluon shadowing according to EPS09LO global fits~\cite{Eps09}. 

\section{\ro photoproduction}

In contrast to \jpsi, \ro photoproduction is hardly sensitive to partonic degrees of freedom and gluon shadowing effects in nuclei. However, measurement of the coherent \ro photoproduction cross section in UPC at LHC, supplemented by STAR results at RHIC energies~\cite{star}, should help to verify \ro photoproduction models which differ by factor 2 in the predicted cross sections~\cite{sl,gm,rho}.

\begin{figure}
\centering
\includegraphics[width=0.49\linewidth]{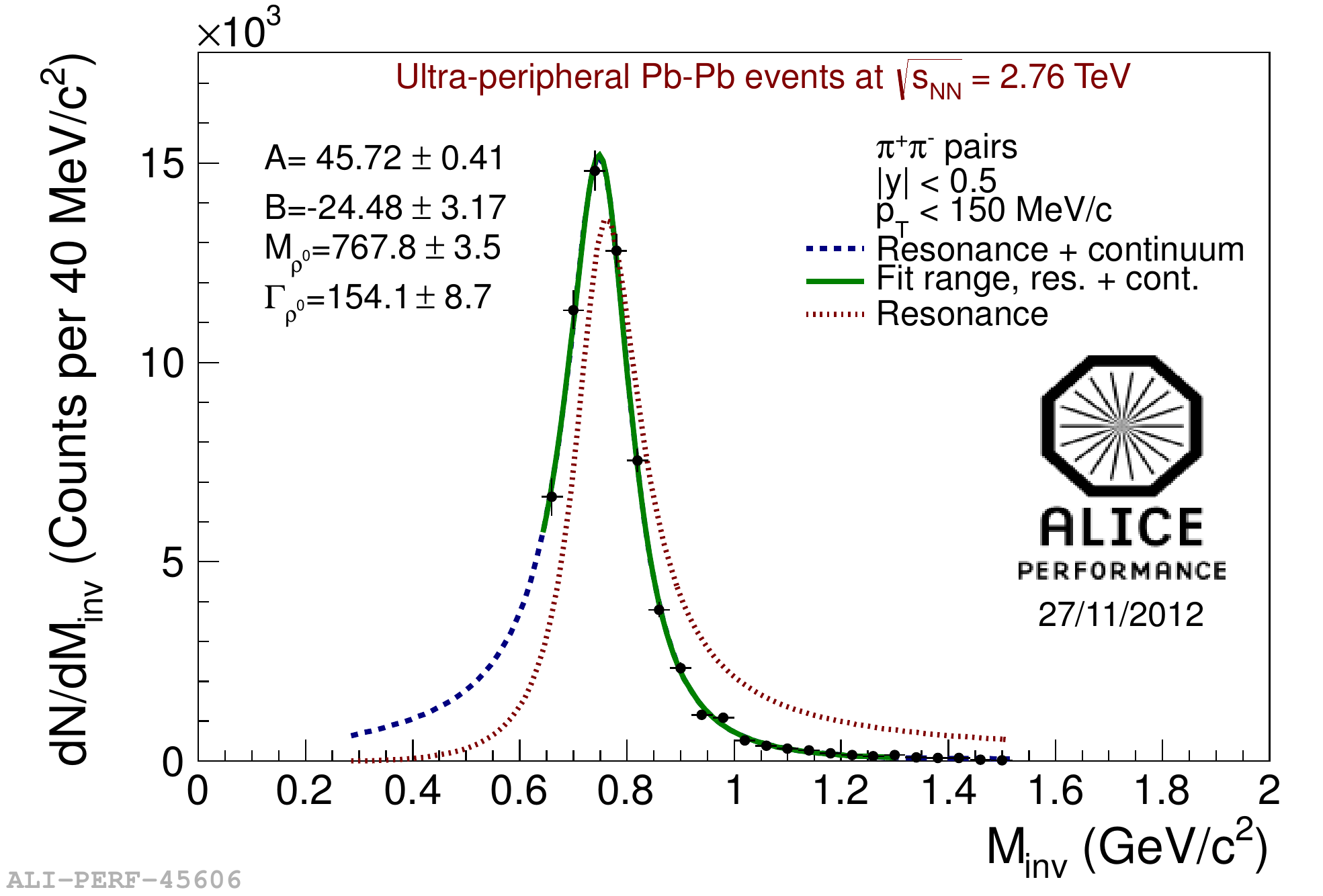}
\caption{Unfolded invariant mass distribution 
for $\pi^+\pi^-$  pairs  in ultraperipheral Pb+Pb collisions.}
\label{fig:rho}
\end{figure}

ALICE measured \ro photoproduction in the $\pi^+\pi^-$ channel based on 2010 Pb+Pb data with the trigger requiring at least two hits in SPD and TOF and vetoes in VZERO counters at forward and backward directions. The unfolded $\pi^+\pi^-$ invariant mass distribution, shown in fig.~\ref{fig:rho}, was fitted with the function which includes variable width Breit-Wigner and non-resonant $\pi^+\pi^-$ contributions. The obtained mass $M_{\rho^0} = 767.8\pm 3.5$ MeV$/c^2$ and the width
$\Gamma_{\rho^0} = 154.1\pm 8.7$ MeV$/c^2$ are compatible with PDG values. In order to extract the coherent \ro yield, $p_{T}$ distribution for the selected $\pi^+\pi^-$ pairs was fitted with Monte-Carlo templates for coherent and incoherent samples from STARLIGHT. The fraction of incoherent events was found to be $\sim7$\% for $p_{\rm T} < 150$ MeV$/c$. The coherent \ro peak appears to be slightly narrower in data than in the STARLIGHT simulation.

\section{Conclusions and outlook}

ALICE made the first measurement on the exclusive \jpsi{} photoproduction in Pb-Pb collisions at LHC both at forward and central rapidity.  The coherent \jpsi{} cross section was found to be in good agreement with those models which include nuclear gluon shadowing consistent with EPS09 parametrization~\cite{Eps09}. 

In the beginning of 2013, ALICE took p+Pb data with dedicated UPC triggers offering a unique opportunity to measure exclusive quarkonium photoproduction on protons. The exclusive \jpsi{} photoproduction, previously studied  in ep collisions at HERA, provides a direct tool to constrain the gluon density in proton down to $x \sim 10^{-4}$~\cite{mnrt}.  ALICE p+Pb UPC data allow to extend HERA results from $W_{\gamma p} \sim 300$ GeV up to TeV scale ($W_{\gamma p}$ is the $\gamma p$ center-of-mass energy) and to study gluon distributions in the wide range of Bjorken $x$ from $10^{-2}$ to $10^{-5}$.

Diffractive particle production in pp collisions at LHC is another interesting subject studied by ALICE and closely related to UPC. Results on inelastic, single and double diffractive cross sections at 0.9, 2.76 and 7 TeV have been recently published by ALICE~\cite{diffraction}. Special effort is also devoted to the studies of central diffraction corresponding to the double-pomeron exchange mechanism and experimentally identified by gaps in the forward and backward directions.

\end{document}